\documentclass[groupedaddress,superscriptaddress,amsmath,amssymb,aps,prl,twocolumn,longbibliography,10pt]{revtex4-2}

\usepackage{svg} 
\usepackage{graphicx}
\usepackage{dcolumn}
\usepackage{bm}
\usepackage[bookmarks=true,colorlinks]{hyperref}

\usepackage{pdfpages}
\usepackage{etoolbox}
\makeatletter
\patchcmd{\@outputpage@head}{\@ifx{\LS@rot\@undefined}{}{\LS@rot}}{}{}{}
\makeatother

\begin{document}

\title{Taming spin susceptibilities in frustrated quantum magnets: \\
Mean-field form and approximate nature of the quantum-to-classical correspondence }

\author{Benedikt Schneider}
\affiliation{Department of Physics and Arnold Sommerfeld Center for Theoretical
Physics, Ludwig-Maximilians-Universit\"at M\"unchen, Theresienstraße~37,
80333 Munich, Germany}
\affiliation{Munich Center for Quantum Science and Technology (MCQST), 80799 Munich, Germany}

\author{Bj\"orn Sbierski}
\affiliation{Institut für Theoretische Physik, Universit\"at T\"ubingen, Auf der Morgenstelle 14, 72076 T\"ubingen, Germany}

\date{\today}

\begin{abstract}
In frustrated magnetism, the empirically found quantum-to-classical correspondence (QCC) matches the real-space static susceptibility pattern of a quantum spin-$1/2$ model with its classical counterpart computed at a certain elevated temperature. This puzzling relation was observed via bold line diagrammatic Monte Carlo simulations in dimensions two and three. The matching was within error bars and seemed valid down to the lowest accessible temperatures $T$ about an order of magnitude smaller than the exchange coupling $J$.  Here, we employ resummed spin diagrammatic perturbation theory to show analytically that the QCC breaks weakly at fourth order in $J/T$ and provide the approximate mapping between classical and quantum temperatures. Our treatment further reveals that QCC is an indication of the surprising accuracy with which static correlators can be approximated by a simple renormalized mean-field form. We illustrate this for all models discussed in the context of QCC so far, including a recent example of the $S=1$ material $\mathrm{K}_2\mathrm{Ni}_2(\mathrm{SO}_4)_3$. The success of the mean-field form is traced back to partial diagrammatic cancellations.
\end{abstract}

\maketitle

\emph{Introduction}.—Frustrated quantum magnets remain at the forefront of current research in condensed matter physics \cite{lacroixIntroductionFrustrated2011}. In this arena, enhanced spin fluctuations might suppress magnetic ordering and conspire to stabilize delicate highly-entangled quantum states characterized by long-range entanglement and fractional excitations at low temperature $T$ \cite{kitaev_anyons_2006,savaryQuantumSpin2016}. But how low is “low”? And which experimental observables reveal the sought-after quantum spin liquid properties unambiguously? 

\begin{figure*}[t]
    \centering
    \includegraphics[width = \textwidth]{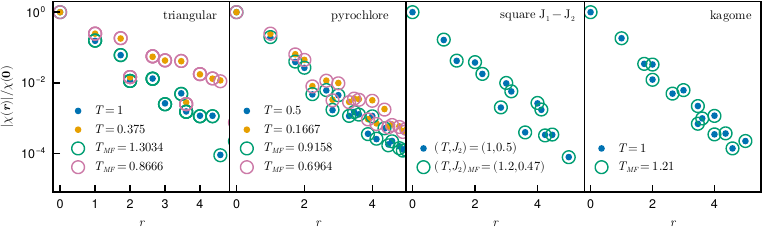}
    \vspace{-0.7cm}
    \caption{Absolute values of normalized real-space spin susceptibility for AFM $S=1/2$ Heisenberg models with $J_1=1$ and temperature $T$ (dots, data taken from BDMC calculations of Refs.~\cite{kulagin_bold_2013-1,huang_spin-ice_2016,wang_quantum--classical_2020}) compared to the renormalized MF form (circles) parameterized with temperature $T_{MF}$ (see text) optimized for best agreement. The almost perfect match for a wide range of models and for rather low $T$ shows the accuracy of the renormalized MF form in $d>1$ dimensions. For pyrochlore, the mean value of $|\chi(\mathbf{r})|$ with the same $r$ is plotted to match the presentation in Fig.~4 of Ref.~\cite{huang_spin-ice_2016}. }
    \label{fig:MF}
\end{figure*}

While complete answers to these questions remain elusive even in theory, impressive progress has been made on the numerical front \cite{sandvik_computational_2010, lohmann_tenth-order_2014, hauschildEfficientNumerical2018, muller2024pseudo,liu2018gapless,chen2018exponential}.
During this endeavor, in 2013, a particularly puzzling empirical observation appeared for the triangular lattice quantum $S=1/2$ Heisenberg anti-ferromagnet (AFM) \cite{kulagin_bold_2013-1}: The bond-resolved spin susceptibility $\chi(\mathbf{r})$ [also known as the static spin correlator, see Eq.~\eqref{eq:chi_Def}] was computed via bold line diagrammatic Monte Carlo (BDMC). For all attainable $T\!\geq \!0.375 J$, the intricate and highly featured normalized pattern $\chi(\mathbf{r})/\chi(\mathbf{0})$  (see Fig.~\ref{fig:MF} left) can be matched by correlation data obtained from the classical ($S=\infty$) vector-spin version of the same model at an empirically obtained elevated temperature $T^{(c)}\!>T$! This was dubbed the quantum-to-classical correspondence (QCC) \footnote{QCC is not to be confused with the fact that quantum phase transitions in $d$ spatial dimensions can share a universality class with a classical counterpart in $d+1$ spatial dimensions \cite{sondhi_continuous_1997}.}.

The BDMC \cite{kulagin_bold_2013} is one of the few numerical methods which remains operative for frustrated quantum spin models in high spatial dimensions ($d=2,3$) and moderately low $T/J \gtrsim 0.1$. It builds on a complex fermionic representation of spin $S=1/2$ operators and stochastically samples millions of skeleton Feynman diagrams \cite{prokof2007bold}. Importantly, results for $\chi(\mathbf{r})$ from BDMC are not exact but come with error bars of $\simeq1\%$. The matching of QCC above is to be understood within these error bars.
\begin{figure}
    \centering
    \includegraphics[width = \columnwidth]{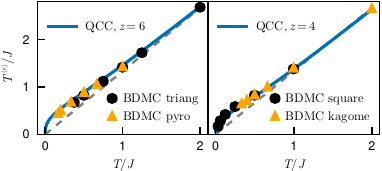}
    \vspace*{-0.8cm}
    \caption{QCC for systems with nearest-neighbor AFM Heisenberg coupling $J$, left for triangular and pyrochlore lattice with coordination number $z=6$ and right for kagome and square lattice ($z=4$). 
    The markers are reproduced from the BDMC studies of Refs.~\cite{kulagin_bold_2013-1,huang_spin-ice_2016,wang_quantum--classical_2020} and indicate the temperature $T^{(c)}$ at which the normalized susceptibility pattern of the classical vector-spin model matches the one of the quantum $S=1/2$ model at temperature $T$ within error bars.
    The solid lines show the analytical prediction for the \emph{approximate} QCC obtained by numerically inverting Eq.~\eqref{eq:QCC}, the dashed line denotes the high-$T$ limit $T^{(c)}=4T/3$.}
    \label{fig:OneParam}
\end{figure}

In the years following its initial observation on the triangular lattice, QCC was found wherever BDMC was aimed at:
The square- and kagome lattice in $d=2$ \cite{kulagin_bold_2013-1,wang_quantum--classical_2020}, the pyrochlore lattice in $d=3$ \cite{huang_spin-ice_2016} as well as $J_1\!-\!J_2$ models on the square and anisotropic triangular lattices in $d=2$ \cite{wang_quantum--classical_2020}. The corresponding $T^{(c)}(T)$ for various nearest-neighbor models is compiled in Fig.~\ref{fig:OneParam} (markers). 
Recently, QCC was also observed in a $S=1$ model for the material $\mathrm{K}_2\mathrm{Ni}_2(\mathrm{SO}_4)_3$ consisting of two interconnected $d=3$ trillium lattices \cite{gonzalez_dynamics_2024}, here the quantum data was obtained from the pseudo-fermion functional renormalization group (PFFRG) \cite{muller2024pseudo}.

As of today, the QCC remains puzzling with overwhelming empirical evidence but no explanation. Why do the celebrated quantum fluctuations maximized for the smallest spin $S\!=\!1/2$ merely seem to be accounted for classically by an effective heating? And would the QCC break down and reveal an approximate nature if a more powerful successor of BDMC could reach lower $T$ and smaller error bars? If not, Ref.~\cite{kulagin_bold_2013-1} speculated, could the $T^{(c)}(T)$ curve be extrapolated to $T=0$ such that a classical simulation would reveal properties of highly sought-after quantum ground states? 

In this letter we shed analytical light on the origin of QCC building on recent progress in spin-$S$-diagrammatics~\cite{schneider2024dipolar}. The QCC turns out to be only approximate for all finite $S$ and dimensions $d$. We quantify QCC's failure at order $[J/T]^4$, provide a closed-form expression of the (approximate) $T^{(c)}(T)$ curve, and reveal a connection between QCC and the surprising success of a renormalized mean-field (MF) ansatz for $\chi(\mathbf{r})$. We also quantitatively explain the failure of QCC for the $d=1$ Heisenberg chain at intermediate temperature, as empirically pointed out already in Ref.~\cite{kulagin_bold_2013-1}.

\begin{figure}
    \centering
    \includegraphics{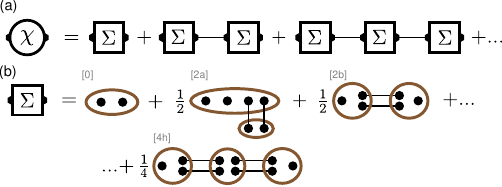}
    \caption{(a) Diagrammatic representation of the Larkin equation \eqref{eq:Larkin}. Lines represent the coupling $ -J$. (b) Some diagrams contributing to $\Sigma$. For details see Ref.~\cite{schneider2024dipolar}. Diagram $[4h]$ and a two-chain $[2b]^2$ contribute to $\Sigma(\mathbf{k})^{-1}$ as $\sim J^4 \gamma(\mathbf{k})^2$. 
    \label{fig:Diagrams}}
\end{figure}

\emph{Perturbation theory}.—We present a resummed perturbative expansion of the susceptibility in spin-$S$ Heisenberg models. For ease of presentation, we here restrict to nearest-neighbor models on $N$-site Bravais lattices with single atomic bases (e.g.~triangular or square lattices),
\begin{align}
    H= J\sum_{\langle\mathbf{r},\mathbf{r'}\rangle}   \mathbf{S}_\mathbf{r} \cdot \mathbf{S}_{\mathbf{r'}},  \label{eq:Hamiltonian}
\end{align}
for generalizations see the End Matter. 
The momentum-space susceptibility or static spin correlator (at zero Matsubara frequency) is
\begin{align}
    \chi(\mathbf{k}) = \frac{1}{N}\sum_{\mathbf{r},\mathbf{r'}} e^{-i \mathbf{k}\cdot(\mathbf{r}-\mathbf{r'} )}\! \int^\beta_0 \!\mathrm{d}\tau \langle S_{\mathbf{r}}^z(\tau)S_{\mathbf{r'}}^z(0) \rangle,
    \label{eq:chi_Def}
\end{align}
where $\beta=1/T$ and $S_{\mathbf{r}}^z(\tau)$ is the Heisenberg spin operator at imaginary time $\tau$. The Larkin equation \cite{vaks_self-consistent_1967,izyumov_statistical_1988} expresses Eq.~\eqref{eq:chi_Def} via the set of one-$J$-irreducible (static) two-legged spin-correlator diagrams $\Sigma(\mathbf{k})$,
\begin{align}
    \chi(\mathbf{k})^{-1} =  \Sigma(\mathbf{k})^{-1} + J \gamma(\mathbf{k}), \label{eq:Larkin}
\end{align}
where $\gamma(\mathbf{k}) = \frac{1}{N}\sum_{\langle\mathbf{r},\mathbf{r'}\rangle} e^{-i \mathbf{k}\cdot(\mathbf{r}-\mathbf{r'} )}$ is the spatial Fourier transform of the real-space coupling pattern normalized to unit strength. A diagrammatic representation of Eq.~\eqref{eq:Larkin} is shown in Fig.~\ref{fig:Diagrams}(a) and a few low-order in $J$ contributions to $\Sigma$ are depicted in Fig.~\ref{fig:Diagrams}(b). 

Replacing the exact $\Sigma$ in Eq.~\eqref{eq:Larkin} with its contribution at order $J^0$, i.e.~with the free-spin (Curie) susceptibility $\Sigma^{(0)}\!=\!\beta b_1$ [where $b_1\!=\!S(S\!+\!1)/3$] yields the MF approximation \cite{izyumov_statistical_1988}. Consequently, terms of higher order capture corrections beyond MF. 
It is convenient to parameterize the \emph{exact} (inverse) susceptibility $\chi(\mathbf{k})^{-1}$ of Eq.~\eqref{eq:Larkin} in terms of a renormalized MF part and a correction $\epsilon(\mathbf{k})$ characterized via its beyond-MF momentum dependence, 
\begin{equation}
\beta[\chi(\mathbf{k})]^{-1} = f + g \gamma(\mathbf{k})  +  \epsilon(\mathbf{k}).\label{eq:full_invchi}
\end{equation}
In MF approximation, $f= 1/b_1$,  $g=\beta J$ and $\epsilon(\mathbf{k}) = 0$.   In that sense the exact $f$ is the  (dimensionless) renormalized on-site inverse susceptibility, while $g$ describes the renormalized coupling. We refer to the right-hand side of Eq.~\eqref{eq:full_invchi} with $\epsilon(\mathbf{k}) \!=\! 0$ as renormalized MF form.

Recent methodological progress \cite{schneider2024dipolar} provides the expansion of $\Sigma$ in the dimensionless coupling $X=\beta J$ complete to $O(X^4)$ for various spin models and arbitrary lattice geometry. Focusing on the static correlator and the Heisenberg case, the right-hand side of Eq.~\eqref{eq:full_invchi} reads
\begin{align}
    f = &  \frac{1}{b_1} +  
   z \left(\frac{1}{6} (6b_1+1)
   X^2+\frac{1}{24} (4b_1+1) X^3\right) \nonumber\\
    & -  \frac{I^{(3)}}{6}b_1 (6b_1+1)   X^3 
      + {O}(X^4),\label{eq:f}\\
   g =& X + \frac{X^2}{12} + \frac{X^3}{120} \left(48
  b_1^2+16b_1+3\right)+{O}(X^4),\label{eq:g}\\
  \epsilon(\mathbf{k}) \!=& -\frac{X^4 b_1}{720}\! \biggl[\gamma^2(\mathbf{k})- z- \frac{I^{(3)}}{z}\gamma(\mathbf{k}) \biggr]  + {O}(X^5).\label{eq:eps}
\end{align}
We defined $I^{(n)} =\int_\mathbf{k} \gamma^n (\mathbf{k})$. The special case $I^{(2)}=z$ counts the neighbors per site (coordination number). For example, the nearest-neighbor Heisenberg model on the $d$-dimensional hyper-cubic lattice is characterized by $\gamma(\mathbf{k})=2\sum_{l=1}^d\cos k_l$, $ z = 2 d$ and $I^{(3)} = 0$. The second and third term in brackets of Eq.~\eqref{eq:eps} subtract contributions $\sim\gamma(\mathbf{k})^2$ that can be associated to $O(X^4)$ terms in $f$ and $g$, respectively.

For all the following arguments, it is crucial that corrections to the renormalized MF form in $\epsilon(\mathbf{k})$ appear only from order $X^4$ on, meaning they are suppressed at high temperatures. Further, the small prefactor in Eq.~\eqref{eq:eps} results from partial diagrammatic cancellation, $\epsilon(\mathbf{k}) \sim (b_1^{-1} [\mathrm{2b}]^2-[\mathrm{4h}])+{O}(X^5)$, see Fig.~\ref{fig:Diagrams}.

\emph{Approximate QCC}.—To address the QCC puzzle \cite{kulagin_bold_2013-1} we consider classical (unit-vector) spins via the $S\rightarrow \infty$ limit. This is straightforward in our general-$S$ formalism after spin-operators in Eqns.~\eqref{eq:Hamiltonian} and \eqref{eq:chi_Def} are rescaled by $1/S$. The resulting classical coupling and susceptibility are denoted with superscript (c), hence $\chi^{(c)}(\mathbf{k})^{-1} = \underset{S\rightarrow\infty}{\mathrm{lim}}  S^2 \chi(\mathbf{k})^{-1}$ with $X \!\xrightarrow{}\! X^{(c)}/S^2$. Note that $b_1$ also depends on $S$, hence the right-hand sides of Eqns.~\eqref{eq:f} to \eqref{eq:eps} simplify in the limit $S\rightarrow \infty$. 

However, as shown above, the susceptibility $\chi(\mathbf{k})$ remains a function of two parameters $f,g$ for all $S$ with $O(X^4)$ corrections. The QCC was discussed for the normalized susceptibility $\chi(\mathbf{k})/\chi(\mathbf{r}=\mathbf{0})$ where $\chi(\mathbf{r}=\mathbf{0})= \int_\mathbf{q} \chi(\mathbf{q})$. The (inverse of the) normalized susceptibility can be computed straightforwardly for both finite $S$ and the classical case. We find an approximate analytic mapping between the two (up to and including order $(X^3)\sim[X^{(c)}]^3$) where $f$ is fixed by the normalization and $g$ can be fixed by relating $X$ to $X^{(c)}$,
\begin{equation}
    X \!=\!\frac{X^{(c)}}{3 b_1} - \frac{(X^{(c)})^2}{108 b_1^2} + (X^{(c)})^3\frac{ 15 b_1 z \!-\! 12 b_1\!-\!1}{2430 b_1^3} + {O}(X^4).\label{eq:QCC}
\end{equation}
Equivalently, this relates the temperature $T/J=1/X$ of the quantum spin-$S$ system to that of the classical system, $T^{(c)}/J=1/X^{(c)}$. In the End Matter we generalize Eq.~\eqref{eq:QCC} to $J_1\!-\!J_2\!-\!\ldots$ Heisenberg models with multiple (equivalent) basis sites. For $J_1$-models on lattices with non-trivial basis (e.g.~kagome or pyrochlore), Eq.~\eqref{eq:QCC} is not changed.
This is our first main result.
 
In Fig.~\ref{fig:OneParam}, we plot the numerical inverse of Eq.~\eqref{eq:QCC} [without the unknown contribution $O(X^4)$] for various $S=1/2$ systems (lines). Note that the lattice only enters via the coordination number $z$. Results are in good agreement with the empirical data points reproduced from the BDMC studies of various $z=4$ and $z=6$ models in Refs.~\cite{kulagin_bold_2013-1,huang_spin-ice_2016,wang_quantum--classical_2020}.

For all lattices considered here, the corresponding classical temperature is always higher than the quantum temperature. Intuitively, quantum fluctuations heat up the system \cite{kulagin_bold_2013-1}. However, this is not generally the case with the spin-dimer as a counter-example, see End Matter.


\emph{Renormalized MF form of susceptibility}.—The QCC is only approximate and breaks down at order $X^4$ where first corrections $\epsilon(\mathbf{k})$ to the renormalized MF form appear, see Eq.~\eqref{eq:eps}. However, this correction vanishes in the classical limit, $\lim_{S\xrightarrow{} \infty} S^2\epsilon(\mathbf{k}) = 0 +{O}((X^{(c)})^6)$.  Therefore, from fourth order on, the momentum dependence of the quantum and classical $\chi(\mathbf{k})$ is inherently different, making it fundamentally impossible to extend the mapping $X(X^{(c)})$ in Eq.~\eqref{eq:QCC} for the full $\chi(\mathbf{k})$ to order $X^4$. The QCC is therefore always of approximate nature and, contrary to speculations \cite{kulagin_bold_2013-1}, fails for $T \rightarrow 0$.

However, it turns out that corrections $\epsilon(\mathbf{k})$ to the renormalized MF form of $[T\chi(\mathbf{k})]^{-1}$ are relatively small even for small $T/J$ (large $X$). This can be anticipated from the partial cancellation of diagrams as mentioned above, but it can also be inferred empirically by existence of the QCC even for very low $T/J$, e.g.~in the pyrochlore case down to $T/J=0.1667$ \cite{huang_spin-ice_2016}. Formally, the renormalized MF form of $[T\chi(\mathbf{k})]^{-1}$ is a good approximation to the exact value as long as the minimum of the first two terms $\Delta=f-g z$ (the MF gap) is large compared to $\epsilon(\mathbf{k})$, such that $\|\epsilon\|_1/\Delta \ll 1$. Here we use a $L_1$ norm in real-space,  $\| \epsilon \|_1 = \sum_{\mathbf{r'}}|\epsilon_{\mathbf{r},\mathbf{r'}}|$.

In the following we show that for the $d=2,3$ models considered  in the context of QCC the susceptibility patterns are very accurately approximated by $\epsilon(\mathbf{k}) \approx 0$ well into the cooperative paramagnetic regime $T/J \gtrsim 0.1$. In Fig.~\ref{fig:MF} we match the $S=1/2$ BDMC data from various models treated in Refs.~\cite{kulagin_bold_2013-1,huang_spin-ice_2016,wang_quantum--classical_2020} to the renormalized MF form with the empirically optimized MF temperature $T_{MF}= \frac{f b_1}{ g} J$. 
For the $J_1\!-\!J_2$ models with the renormalized MF form $[T\chi(\mathbf{k})]^{-1}\simeq f+g_1 \gamma_1(\mathbf{k})+g_2 \gamma_2(\mathbf{k})$ (c.f.~End Matter), also $J_2$ is adjusted, see the labels in Fig.~\ref{fig:MF}. In all instances, the renormalized MF and BDMC data fit very well, even at temperatures below which the analytical mapping based on the truncated version of Eq.~\eqref{eq:QCC} would yield reasonable results. This is our second main result. Importantly, this goes beyond the empirical QCC \cite{kulagin_bold_2013-1} since the exact classical susceptibilities, to which the quantum susceptibilities were matched so far, also feature $O([X^{(c)}]^6)$ corrections to the renormalized MF form, c.f.~the discussion above.

\begin{figure}[t]
    \centering
    \includegraphics[width = \columnwidth]{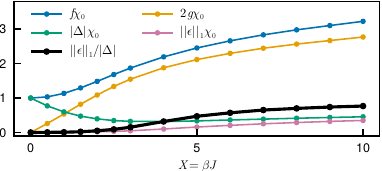}
    \vspace*{-0.6cm}
    \caption{Parameters of the inverse susceptibility in Eq.~\eqref{eq:full_invchi}  for the cyclic $S=1/2$ AFM Heisenberg chain of length $L=128$. Lines are guides to the eye. For better visibility $f,g,\epsilon$ are multiplied by $\chi_0\equiv\chi(r=0)$.  Breakdown of the renormalized MF form (and of QCC) is rooted in a non-negligible $\| \epsilon \|_1/\Delta$.}
    \label{fig:1DHeisenberg}
\end{figure}

As stated already by Kulagin et al.~\cite{kulagin_bold_2013-1} the QCC clearly fails in the $d=1$ AFM Heisenberg chain. The susceptibility of this model's classical counterpart is exactly described by the renormalized MF form with $f = \frac{3 u^2+3}{1-u^2}$, $g = \frac{3 u}{1-u^2}$ and $\epsilon(\mathbf{k})=0$ where $u =\coth (X) - 1/X$ \cite{fisher1964magnetism}:  
Therefore, QCC can be observed as long as $\|\epsilon\|_1  /\Delta \ll 1$ for the quantum correlator. We calculate the susceptibility for the $S=1/2$ model with quantum MC \cite{PolletWorm,PolletWormCode}. 
In Fig.~\ref{fig:1DHeisenberg} we show that already for $T/J = 0.33$, the corrections to renormalized MF form are sizable,  $\|\epsilon\|_1 / \Delta \approx 15\% $. This 
means that the range of $T/J$ where QCC could be observed is significantly smaller in $d=1$ than in the $d=2,3$ models, where no violation was found at the temperatures available to BDMC. This empirical finding can be rationalized in a $1/d$-expansion \cite{schneider2024dipolar} where corrections to the renormalized MF form at the MF-critical temperature scale as $\epsilon(\mathbf{k}) \sim \frac{1}{d^2}$.
\begin{figure*}[t!]
    \centering
    \includegraphics[width = \textwidth]{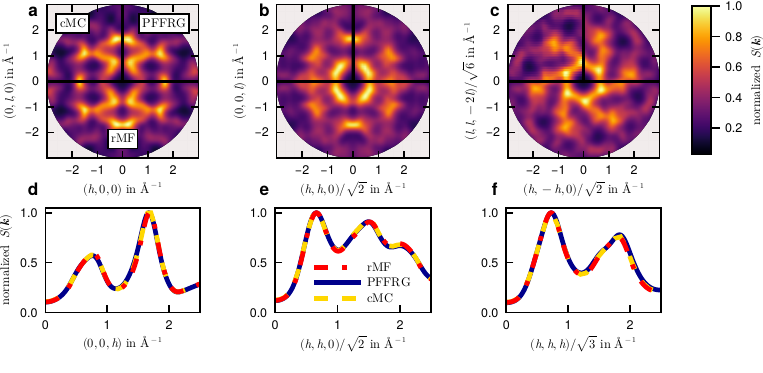}
    \vspace*{-1.2cm}
    \caption{Static spin structure factor $S(\mathbf{k}) = \sum_{\nu,{\nu^\prime}}\chi_{\nu,{\nu^\prime}}(\mathbf{k}) f_{\mathrm{Ni}^{2+}}(\mathbf{k})$  of $\mathrm{K}_2\mathrm{Ni}_2(\mathrm{SO}_4)_3$ along different planes in reciprocal space, where  $f_{\mathrm{Ni}^{2+}}(\mathbf{k})$ is the form factor of the  $\mathrm{Ni}^{2+}$ ions. Panels (a)-(c) show classical MC results at $T = 0.35 J_4$ (top left part of each panel), PFFRG results at flow parameter $\Lambda = 0.58 J_4$ (top right), and a fit to the renormalized MF form (rMF, lower half) at $T_{\mathrm{MF}} = 0.787 J_4$. Panels (d)-(f) display line cuts along three principal directions. The classical MC and PFFRG data were taken from Ref.~\cite{gonzalez_dynamics_2024}.}
    \label{fig:Trillium}
\end{figure*}

Finally, we test the renormalized MF form for a model of a complex realistic material $\mathrm{K}_2\mathrm{Ni}_2(\mathrm{SO}_4)_3$ where QCC was empirically observed by Gonzalez et al.~\cite{gonzalez_dynamics_2024}.
According to  \emph{ab-initio} calculations \cite{gonzalez_dynamics_2024}, this $d\!=\!3$ material realizes a $S\!=\!1$ Heisenberg model on two interconnected trillium lattices with $\{J_1,J_2,J_3,J_4,J_5\} = \{0.066, -0.026, 0.144, 1 , 0.479\}$ with dominant $J_4=1$. 
This model was treated with the ground state PFFRG \cite{muller2024pseudo} with results extracted 
at finite Matsubara frequency cutoff parameter $\Lambda = 0.58 J_4$. The latter was argued to act (at least qualitatively) as a finite temperature $T\sim \Lambda$ \cite{iqbal_quantum_2019}.
Unlike BDMC, the error of PFFRG is uncontrolled. However, empirical confirmation of QCC with $T^{(c)} = 0.35 J_4$ found in Ref.~\cite{gonzalez_dynamics_2024} suggests that the susceptibility is well described by a renormalized MF form. In Fig.~\ref{fig:Trillium} we match the classical MC and PFFRG susceptibilities \cite{gonzalez_dynamics_2024} along multiple planes and cuts in $\mathbf{k}$-space to a renormalized MF form at $T_{\mathrm{MF}} = 0.787 J_4$ and all $J_m$ unchanged. The quality could be further improved by also renormalizing the latter.

\emph{Conclusion}.—The approximate description of quantum spin systems with classical theories is a recurring theme \cite{chakravartyTwodimensionalQuantum1989,sachdev_quantum_2011,dahlbomQuantumtoclassicalCrossover2024,park2024quantum} and cancellations of quantum effects have been observed for various specific spin models on particular lattices \cite{wuTunnelinginducedRestoration2019,wangRenormalizedClassical2024}. However, the so far empirical QCC for the spin susceptibility on general two- and three dimensional Heisenberg models \cite{kulagin_bold_2013-1} revisited in this letter stands out by its surprising universality and accuracy even for low temperatures $T$ down to about an order of magnitude below $J$. Using resummed spin diagrammatic perturbation theory, we showed rigorously that the QCC is only approximate and fails at order $(J/T)^4$. At this order, the spatial dependence of the exact susceptibility for $S<\infty$ deviates from a simple renormalized MF form.

In this sense, the QCC can be understood as a symptom of a more fundamental insight put forward in this work: The renormalized MF form almost perfectly accounts for a plethora of spin susceptibility patterns in $d=2$ and $d=3$ reported in the literature on the basis of numerically expensive computations. We rationalize this from the partial diagrammatic cancellations of corrections as explicitly observed at order $(J/T)^4$ but likely operative also at higher orders.

The success of the renormalized MF form is surprising, since ordinary MF theory is only  valid for high temperatures or lattices with large coordination numbers.
For future work, it is thus an interesting question if the renormalized MF ansatz could inform the development of novel theoretical (renormalization-)methods or experimental data analysis. The latter could be especially relevant for atom-tweezer array quantum simulators where $\chi(\mathbf{r})$ should be directly accessibly due to single-site control and measurement capabilities \cite{manovitzQuantumCoarsening2024}. As a first step, it would be useful to extend the analytical (and approximate) QCC mapping of Eq.~\eqref{eq:QCC} to higher orders in $(J/T)$ for better estimation of the renormalized MF parameters.

\begin{acknowledgements}
\emph{Acknowledgments}.—We thank Mat\'ias G.~Gonzalez, Philip Osterholz, Johannes Reuther, Nepomuk Ritz,  Lode Pollet and Manuel Weber for helpful discussions.
We acknowledge support from DFG grant no.~524270816. B.Sb.~acknowledges support from DFG through the Research Unit FOR 5413/1 (grant no.~465199066).  B.Sch.~acknowledges funding from the Munich Quantum Valley, supported by the
Bavarian state government with funds from the Hightech Agenda Bayern Plus.
\end{acknowledgements}

\bibliography{QCC}

\appendix
\onecolumngrid
\begin{center}
    \textbf{\large End Matter}
\end{center}
\twocolumngrid

\emph{Multi-parameter models}.—We generalize the QCC mapping for finite-range multi-parameter $J_1-J_2-...-J_M$ models on arbitrary lattices. However, we restrict to the case where all lattice sites are equivalent by symmetry so that $\chi(\mathbf{r}=\mathbf{0})$ used for normalization in the QCC is unambiguous. We start with a spin-$S$ Heisenberg Hamiltonian of the type
\begin{align}
    H=\frac{1}{2}\sum_{\mathbf{r_{\nu}},\mathbf{r^\prime_{\nu^\prime}}}  J_{(\mathbf{r_{\nu}},\mathbf{r^\prime_{\nu^\prime}})}  \mathbf{S}_\mathbf{r_{\nu}} \cdot \mathbf{S}_{\mathbf{r^\prime_{\nu^\prime}}}  \label{eq:App_Hamiltonian}
\end{align}
and on a general Bravais lattice spanned by basis vectors $\{ \mathbf{e}_1,\ldots ,\mathbf{e}_n\}$ with a $\mu$-atomic basis $\{ \boldsymbol{\delta}_1,\ldots, \boldsymbol{\delta}_\mu\}$,  such that each spin position is uniquely described by $\mathbf{r_{\nu}}=\sum_{i} r_i\mathbf{e}_i + \boldsymbol{\delta}_\nu$ with $r_{i}\in\mathbb{Z}$ and $\nu=1,2,...,\mu$.
We define the matrix-valued susceptibility (or static correlator, at vanishing Matsubara frequency) 
\begin{align}
    \chi_{\nu\nu^\prime}(\mathbf{k}) = \frac{1}{N} \!\! \sum_{\mathbf{r},\mathbf{r^\prime}} e^{-i \mathbf{k}\cdot(\mathbf{r_{\nu}}-\mathbf{r^\prime_{\nu^\prime}} )}\!\! \int^\beta_0 \!\mathrm{d}\tau \langle S_{\mathbf{r_{\nu}}}^z(\tau)S_{\mathbf{r^\prime_{\nu^\prime}}}^z(0) \rangle.
\end{align}

\emph{Parameterization of the coupling matrix}.—For each coupling parameter $J_{1},J_{2},...,J_{M}$, we define a corresponding unit-strength pattern or coupling matrix 
\begin{equation}
\gamma_{m,\nu\nu^{\prime}}(\mathbf{r},\mathbf{r}^{\prime})=\begin{cases}
1 & :J_{(\mathbf{r}_{\nu},\mathbf{r}_{\nu^{\prime}}^{\prime})}=J_{m},\\
0 & :\mathrm{otherwise},
\end{cases}
\end{equation}
which takes bonds $(\mathbf{r}_{\nu},\mathbf{r}_{\nu^{\prime}}^{\prime})$ as input and outputs unity only if this bond has coupling strength $J_{m}$ and zero otherwise. With this definition, we obtain with $X_m=\beta J_m$ for $m=1,2,...,M$,
\begin{equation}
    \beta J_{(\mathbf{r}_{\nu},\mathbf{r}_{\nu^{\prime}}^{\prime})}=\sum_{m=1}^{M}X_{m}\gamma_{m,\nu\nu^{\prime}}(\mathbf{r},\mathbf{r}^{\prime}).
\end{equation}
Fourier transforming the expression gives
\begin{align}
    \beta \mathbf{J}(\mathbf{k}) = {\beta}\sum_{\mathbf{r},\mathbf{r^\prime}} J_{(\mathbf{r_{\nu}},\mathbf{r^\prime_{\nu^\prime}})} e^{-i \mathbf{k}\cdot(\mathbf{r_\nu}-\mathbf{r^\prime_{\nu^\prime}} )} = \sum_{m=1}^M X_m \boldsymbol{\gamma}_{m}(\mathbf{k}) \label{eq:App_InteractionDecomposition}
\end{align}
where we indicated matrices in sublattice space with bold symbols. The $\boldsymbol{\gamma}_m$ are normalized such that $\boldsymbol{\gamma}_{m}(\mathbf{k}) \delta_{m,n}= \int_{\mathbf{q}} \boldsymbol{\gamma}_{m}(\mathbf{q}) *\boldsymbol{\gamma}_{n}(\mathbf{k}-\mathbf{q})$  where $\int_\mathbf{k}  = \frac{1}{V_{BZ}} \int \mathrm{d} \mathbf{k} $ and $*$ is the Hadamard product in sublattice space. 
Since we assumed that all sites of the lattice are equivalent, the matrices $\gamma_m(\mathbf{k})$  commute $[\boldsymbol{\gamma}_{m}(\mathbf{k}),\boldsymbol{\gamma}_{n}(\mathbf{k})]=0$.

The coordination number $z_m$ with respect to coupling $J_m$ can be expressed via $\int_\mathbf{k} \mathrm{Tr}[\boldsymbol{\gamma}_n (\mathbf{k}) \cdot \boldsymbol{\gamma}_m (\mathbf{k})] =\mu \, z_{m} \delta_{m,n}$
and the number of three-loops made from couplings ${l,m,n}$ is $\mu \, I^{(3)}_{l,m,n}  = \int_{\mathbf{k}} \mathrm{Tr}[\boldsymbol{\gamma}_l (\mathbf{k}) \cdot \boldsymbol{\gamma}_m (\mathbf{k}) \cdot \boldsymbol{\gamma}_n (\mathbf{k})]$ respectively. Recall that $\mu$ is the number of sublattices. These relations straightforwardly follow from the equivalent expressions in real-space. \\

\emph{Examples for coupling matrices}.—As an example, the $J_1-$Heisenberg model on the kagome lattice can be described by $\{ \mathbf{e}_1,\mathbf{e}_2\} = \{ (2,0),(1,\sqrt{3})\}$ and $\{ \boldsymbol{\delta}_1,\boldsymbol{\delta}_2, \boldsymbol{\delta}_3\} = \{(1,0),(\frac{1}{2},\frac{\sqrt{3}}{2}),(-\frac{1}{2},\frac{\sqrt{3}}{2})\}$ where the triangular Bravais lattice sites are in the center of the hexagons. This lattice yields
\begin{align}
\boldsymbol{\gamma}(\mathbf{k}) \!=\!
\resizebox{0.42\textwidth}{!}{$
\left(
\begin{array}{ccc}
 0 & 2 \cos \left(\frac{k_1}{2}\!+\!\frac{\sqrt{3} k_2}{2}\right) & 2 \cos \left(\frac{k_1}{2}\!-\!\frac{\sqrt{3} k_2}{2}\right) \\
 2 \cos \left(\frac{k_1}{2}\!+\!\frac{\sqrt{3} k_2}{2}\right) & 0 & 2 \cos (k_1) \\
 2 \cos \left(\frac{k_1}{2}\!-\!\frac{\sqrt{3} k_2}{2}\right) & 2 \cos (k_1) & 0 \\
\end{array}
\right).$}
\end{align}

For the anisotropic $J_1-J_2-$triangular lattice with horizontal chains coupled by $J_1$ and interchain coupling $J_2$ we have $\{ \mathbf{e}_1,\mathbf{e}_2\} = \{ (1,0),(1,\sqrt{3})\}$ and $ \boldsymbol{\delta}_1 = (0,0)$ which leads to
\begin{align}
  \boldsymbol{\gamma}_1(\mathbf{k}) &= 2 \cos (k_1),\\
   \boldsymbol{\gamma}_2(\mathbf{k}) &= 2 \cos \left(\frac{k_1}{2}+\frac{\sqrt{3} k_2}{2}\right) + 2 \cos \left(\frac{k_1}{2}-\frac{\sqrt{3} k_2}{2}\right).
\end{align}

\emph{Perturbative calculation of $\chi$}.—In matrix form the Larkin equation \eqref{eq:Larkin} reads
\begin{align}
    \boldsymbol{\chi}(\mathbf{k})^{-1} =  \boldsymbol{\Sigma}(\mathbf{k})^{-1} +  \mathbf{J}(\mathbf{k}). \label{eq:App_Larkin}
\end{align}MF theory amounts to using only the $O([\beta J]^0)$ term, $\boldsymbol{\Sigma}^{(0)}(\mathbf{k})= b_1 \delta_{\nu^\prime,\nu}$ with $b_1=S(S+1)/3$ the first derivative of the Brillouin function at zero field.
Combining Eqns.~\eqref{eq:App_Larkin} and \eqref{eq:App_InteractionDecomposition}, we parameterize the \emph{exact} inverse correlator in analogy to the one-parameter case \eqref{eq:full_invchi} with 
\begin{align}
[T\boldsymbol{\chi}(\mathbf{k})]^{-1} = f+\sum_m g_m\boldsymbol{\gamma}_m(\mathbf{k})  +  \boldsymbol{\epsilon}(\mathbf{k}). \label{eq:App_full_invchi}
\end{align}
In MF approximation, $f= \frac{1}{b_1}$,  $g_m= X_m$ and $\boldsymbol{\epsilon}(\mathbf{k}) = 0$.   Analogously to the one parameter case, in the full theory  $f$ is the on-site inverse susceptibility, while $g_m$ describes the renormalized MF coupling parameters and $\boldsymbol{\epsilon}(\mathbf{k})$ collects all corrections to the MF form of the correlator.

Using the theory developed in Ref.~\cite{schneider2024dipolar}, the inverse static correlator of Eq.~\eqref{eq:App_full_invchi} can be expanded in $X_m$ as
\begin{widetext}
\begin{align}
    &f =  \frac{1}{b_1} + \sum _{m}
   z_m \left(\frac{1}{6} (6b_1+1)
   X_m^2+\frac{1}{24} (4b_1+1) X_m^3\right) -\frac{1}{6}b_1 (6b_1+1) \sum _{l,m,n} X_m 
   X_l X_{n} I^{(3)}_{l,m,n}  +{O}(X^4),\\
  & g_m = X_m + \frac{X_m^2}{12} + \frac{1}{120} \left(48
  b_1^2+16b_1+3\right) X_m^3 +{O}(X^4),\\
  &\boldsymbol{\epsilon}(\mathbf{k}) = -\frac{b_1}{720}  \sum_{m,n} (X_{m})^2(X_{n})^2 \left(\boldsymbol{\gamma}_m(\mathbf{k}) \cdot \boldsymbol{\gamma}_n(\mathbf{k}) - \delta_{m,n} z_n - \sum_{l} \frac{I^{(3)}_{l,m,n}}{z_l}\boldsymbol{\gamma}_{l}(\mathbf{k})\right) + {O}(X^5).
\end{align}
\end{widetext}

\emph{Analytic calculation of the approximate QCC}.—Just as in the one-parameter case  $\boldsymbol{\chi}(\mathbf{k})^{-1}$ is of renormalized MF form up to third order in $X$. It is a function of the $M+1$ parameters ${f,g_1,g_2,...,g_M}$. Therefore, when comparing the inverse correlator $\boldsymbol{\chi}(\mathbf{k})^{-1} \chi(\mathbf{r}=\mathbf{0})$ normalized with the local correlator $\chi(\mathbf{r}=\mathbf{0})=\frac{1}{\mu}\mathrm{Tr}\int_{\mathbf{q}}\boldsymbol{\chi}(\mathbf{q})$, with its classical counterpart, we can find a mapping between the two. $f$ is fixed by the normalization and  $g_m$ can be fixed by relating $X_m$ and $X_m^{(c)}$,
\begin{align}
    X_{m} =&\frac{X^{(c)}_{m}}{3 b_1} -\frac{(X^{(c)}_{m})^2}{108 b_1^2} 
      +\left(X_{m}^{(c)}\right)^{3}\frac{15b_{1}z_{m}-12b_{1}-1}{2430b_{1}^{3}} \nonumber\\
      &+\frac{X_m^{(c)}}{162 b_1^2}\sum_{n\neq m}z_{n}\left(X_{n}^{(c)}\right)^{2}
      +{O}(X^4).\label{eq:App_QCC}
\end{align}
This generalizes Eq.~\eqref{eq:QCC} of the main text. As explained there, the mapping is only approximate and ceases to exist rigorously at fourth order in $X$. 

\begin{figure}
    \centering
    \includegraphics{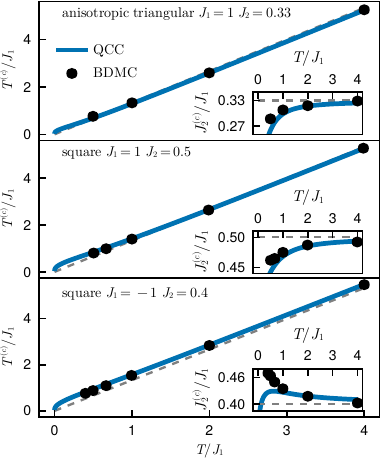}
    \caption{Mapping between the classical and quantum temperatures for two-parameter models studied by BDMC in Ref.~\cite{wang_quantum--classical_2020}. The solid lines are obtained by inverting Eq.~\eqref{eq:App_QCC}. The markers give the same mapping empirically obtained in Ref.~\cite{wang_quantum--classical_2020}. The dashed line indicates the high temperature asymptotic.}
    \label{fig:TwoParam}
\end{figure}

By inverting Eq.~\eqref{eq:App_QCC} we capture the QCC of the two-parameter $J_1\!-\!J_2$ models studied in Ref.~\cite{wang_quantum--classical_2020}. In analogy to Fig.~\ref{fig:OneParam} we compare the empirical data \cite{wang_quantum--classical_2020} with Eq.~\eqref{eq:App_QCC}, see Fig.~\ref{fig:TwoParam}. The classical temperatures $T^{(c)}$ (from $X_1^{(c)}$) and $J_2^{(c)}$ (from $X_2^{(c)}$) are predicted quite well. Only the $J^{(c)}_2$ couplings of the square lattice $J_1-J_2$ with ferromagnetic $J_1$ show a wrong curvature for small quantum temperatures $T/J_1 \lesssim 0.5$. This signals the breakdown of third order perturbation theory.

\emph{QCC for the dimer}.—For $J_1$-Heisenberg models, QCC generally depends on the correlator being a function of two parameters. 
Since the Heisenberg dimer $H=J \mathbf{S}_1 \cdot \mathbf{S}_2$ has only two sites and thus just two susceptibilities (local and non-local), there exists a trivial, exact QCC. 
The classical correlator is given by 
\begin{align}
   T \chi^{(c)}(k) = \frac{1}{3} + \frac{1}{3}(\coth (X^{(c)}) - \frac{1}{X^{(c)}}) \cos (k),
\end{align} 
where $k \in \{0,\pi \}$ and the quantum correlator reads 
\begin{align}
   T \chi(k) =  \frac{e^X-1+X}{2(e^X+3)} + \frac{e^X-1-X}{2(e^X+3)}\cos (k).
\end{align}
Therefore, the exact QCC between the two can be found by solving 
\begin{align}
    (\coth (X^{(c)}) - \frac{1}{X^{(c)}}) = \frac{e^X-1-X}{e^X-1+X}
\end{align}
Setting $J=1$, for large temperatures  $T^{(c)} \simeq \frac{4}{3} T>T$ the system is heated up by the quantum fluctuation but for low temperatures the classical temperature is exponentially reduced  $T^{(c)} =  \left(\frac{2}{T^2} e^{-1/T}\right) T $. This is related to the fact that the the quantum dimer is gapped so that susceptibilities decrease to zero at low temperature where the state of the classical system can still be easily perturbed. To make up for this difference, $T^{(c)}<T$ at low temperatures. Other (trivial) examples where the QCC holds exactly are the cyclic spin trimer, and the Heisenberg model in infinite dimensions.

\end{document}